\documentclass[reprint,amssymb,amsmath,aip,cha,twocolumn, superscriptaddress,frontmatterverbose]{revtex4-2}
\usepackage{graphicx}
\usepackage{dcolumn}
\usepackage{bm}
\usepackage{url,hyperref,lineno}
\usepackage{color}

\begin{document}
\title{Control of seizure-like dynamics in neuronal populations with excitability adaptation related to ketogenic diet}

\author{Sebastian Eydam}
\email{richard.eydam@riken.jp}
\affiliation{Neural Circuits and Computations Unit, RIKEN Center for Brain Science, 2-1 Hirosawa, 351-019 Wako, Japan}

\author{Igor Franovi\'c}
\email{franovic@ipb.ac.rs}
\affiliation{Scientific  Computing  Laboratory,  Center  for  the  Study  of  Complex  Systems, Institute  of  Physics  Belgrade,  University  of  Belgrade,  Pregrevica  118,  11080  Belgrade,  Serbia}

\author{Louis Kang}
\email{louis.kang@riken.jp}
\affiliation{Neural Circuits and Computations Unit, RIKEN Center for Brain Science, 2-1 Hirosawa, 351-019 Wako, Japan}

\date{\today}

\begin{abstract}
We consider a heterogeneous, globally coupled population of excitatory quadratic integrate-and-fire neurons with excitability adaptation due to a metabolic feedback associated with ketogenic diet, a form of therapy for epilepsy. Bifurcation analysis of a three-dimensional mean-field system derived in the framework of next-generation neural mass models allows us to explain the scenarios and suggest control strategies for the transitions between the neurophysiologically desired asynchronous states and the synchronous, seizure-like states featuring collective oscillations. We reveal two qualitatively different scenarios for the onset of synchrony. For weaker couplings, a bistability region between the lower- and the higher-activity asynchronous states unfolds from the cusp point, and the collective oscillations emerge via a supercritical Hopf bifurcation. For stronger couplings, one finds seven co-dimension two bifurcation points, including pairs of Bogdanov-Takens and generalized Hopf points, such that both lower- and higher-activity asynchronous states undergo transitions to collective oscillations, with hysteresis and jump-like behavior observed in vicinity of subcritical Hopf bifurcations. We demonstrate three control mechanisms for switching between asynchronous and synchronous states, involving parametric perturbation of the adenosine triphosphate (ATP) production rate, external stimulation currents, or pulse-like ATP shocks, and indicate a potential therapeutic advantage of hysteretic scenarios.   
\end{abstract}
    
\maketitle

\begin{quotation}
Adaptation has a profound impact on the dynamics of coupled systems across a wide variety of fields. In neuronal systems, 
it has two facets: one associated with synaptic plasticity, and the other related to modification of local kinetics, such as the spiking frequency or the excitability feature. Here we introduce a model of excitability adaptation reflecting the changes in neuronal metabolism linked to the ketogenic diet, a dietary treatment for epilepsy based on replacing carbohydrates with fat. The diet induces a shift 
in the main mechanism of ATP production, which in turn leads to activation of ATP-gated potassium channels in neuronal membranes. We build a model of a heterogeneous, globally coupled population of inherently excitable or tonic spiking excitatory quadratic integrate-and-fire neurons influenced by an ATP-dependent hyperpolarizing current, endowed by an additional equation describing the changes in the total ATP concentration. The stability of solutions and their bifurcations are investigated by analyzing the corresponding mean-field model via the numerical path-following technique. We disclose qualitatively different scenarios of multistability and bifurcations depending on the coupling strength. Interpreting the low-activity asynchronous stationary state as the physiologically desired (homeostatic) one, and the emergence of synchronous solutions featuring collective oscillations as the signature of seizure-like events, we demonstrate three control strategies to suppress the seizure dynamics by inducing a switch to the homeostatic state. Interestingly, such switches may be triggered by the changes in the ATP production rate, as a form of parametric perturbation, both for smooth or hysteretic transitions between synchronous and asynchronous states.
\end{quotation}

\section{Introduction} \label{sec:intro}

The recent advent of the next-generation neural mass (NGNM) models \cite{Montbrio2015,Bick2020,Coombes2019,Byrne2020,Coombes2023} has allowed for a deeper understanding of multistability and critical transitions in the dynamics of neuronal populations, and has provided us with the means to systematically explore the control strategies for switching between the states interpreted in terms of neurophysiology as regular or pathological ones. Earlier phenomenological neural mass models \cite{Coombes2014,Deco2008,Deschle2021} described the dynamics of neuronal populations in terms of firing rate equations, typically assuming a sigmoid-like transfer function between the mean membrane potential and the mean firing rate. Such models relied on the assumption of uncorrelated local dynamics and could neither capture the scenarios where the state of synchrony undergoes qualitative change nor establish an exact relationship between the features of local and collective dynamics. These issues have been resolved in the NGNM models, based on applying the Ott-Antonsen \cite{Ott2008,Ott2009,Luke2013} or the equivalent Lorentzian Ansatz \cite{Montbrio2015} to populations of theta neurons and quadratic integrate-and-fire neurons, respectively. NGNM models characterize population dynamics in terms of mean-field equations for the mean firing rate and the mean membrane potential. While the notion of spike synchrony, classically invoked in neuroscience, is reflected in oscillations of the mean firing rate \cite{Devalle2017}, the NGNM framework further establishes a relation between the mean firing rate and the population synchrony described by the complex Kuramoto order parameter, a synchrony measure inherited from the theory of coupled phase oscillators \cite{Montbrio2015,Byrne2020}.
NGNM models have the advantage of being exact in the thermodynamic limit and are amenable to bifurcation analysis. So far, NGNM models have been used to address a broad range of theoretical problems, including understanding the impact and interplay of 
chemical and electrical synapses \cite{Ratas2016,Pietras2019,Montbrio2020}, the effects of deterministic or stochastic drive \cite{Goldobin2021c,Volo2022,Clusella2022,Pietras2023,
Pyragas2023}, presence of quenched randomness in network topology \cite{Goldobin2021}, finite-size fluctuations \cite{Klinshov2022,Klinshov2023}, interaction of neuronal populations \cite{Ratas2017,Schmidt2018,Pyragas2021}, as well as formation of bumps and waves in neural fields \cite{Byrne2019b,Schmidt2020,Byrne2022}. Moreover, a wide spectrum of applications has been considered, from explaining the onset of gamma and theta-nested gamma oscillations \cite{Devalle2017,Dumont2019,Bi2020,Segneri2020,Keeley2019}, cross-frequency couplings \cite{Ceni2020}, abnormal beta rebound in schizophrenia \cite{Byrne2019}, the impact of deep brain stimulation \cite{Weerasinghe2019}, to studies on working memory \cite{Schmidt2018,Taher2020}, propagation of epileptic seizures \cite{Gerster2021}, and modeling of 
the entire brain functional networks \cite{Byrne2020}. 

A corpus of problems that have so far been less explored within the formalism of NGNM models concerns the impact of neuronal adaptation \cite{Gast2020,Taher2020,Gast2021,Ferrara2023}. In general, neuronal systems feature two broad types of adaptive behavior \cite{Sawicki2023}. The first is synaptic plasticity \cite{Abbott2000,Caporale2008}, which involves modification of coupling strengths due to feedback with neuronal activity. The other type comprises adaptation of local kinetics, including spike frequency adaptation \cite{Ha2017,Fuhrmann2002,Ferrara2023} or changes in neuronal excitability due to metabolic constraints \cite{Katsu2017,Fardet2020,Franovic2022}. Our study 
addresses the latter problem. In particular, we 
use the NGNM framework to investigate how the population dynamics is affected by the excitability adaptation due to a severe change of neuronal energy metabolism triggered by the ketogenic diet (KD) \cite{Lutas2013,Wheless2008,Fei2020,Rho2017,Meira2019,
Rashidy2023}. Constructing the bifurcation diagrams to reveal critical transitions and regions of multistability, we demonstrate different control strategies that may be used to induce switches between asynchronous ("homeostatic”, i.e. neurophysiologically desired equilibrium states with low firing rates) and synchronous, seizure-like regimes.

Within recent years, there has been a growing interest in the impact of energy metabolism constraints, both on maintaining neuronal homeostasis and the emergence of neurological disorders \cite{Roberts2014,Virkar2016,Kroma2021,Franovic2022}. The neuronal energy demand is typically high even in the resting state because it requires maintaining ion fluxes against concentration and electric gradients \cite{Fardet2020}. Such ion fluxes are mainly perpetuated by the $Na^+/K^+$ pump which consumes energy in the form of ATP \cite{Glynn2002,Forrest2014}, produced in the glycolysis process unfolding close to the neuronal membrane. It has long been understood that some neurological disorders, such as epilepsy \cite{Katsu2017,Bazz2017,Patel2018,Kovacs2018} or Parkinson’s disease \cite{Bueller2009,Haddad2015}, involve a strong metabolic component. Epilepsy is the third most prevalent neurological disorder \cite{Beghi2019,Picot2008}, and is characterized by recurring seizures. Despite the advanced anti-convulsive medications, about a third of patients still remain drug-resistant and require alternative therapies \cite{Picot2008}. Among the latter, KD \cite{Lutas2013,Fei2020,Rho2017,Meira2019,Rashidy2023}, whose origins date back to the 1920s, is nowadays often recommended to children with pharmacology-resistant seizures \cite{Neal2010}. KD is based on a 4:1 intake ratio of fat 
vs. carbohydrates plus proteins, which gives rise to a drastic metabolic change called ketosis. In the state of ketosis, the glycolytic ATP production reduces, becoming effectively replaced by an alternative oxidation mechanism in mitochondria, fueled by the ketone bodies generated in the liver using diet-related fatty acids. The KD's effectiveness apparently derives from leveraging anti-convulsive mechanisms that are not (sufficiently) targeted by medications \cite{Lutas2013}. 

\begin{figure}
  \centering
    \includegraphics[scale=0.22]{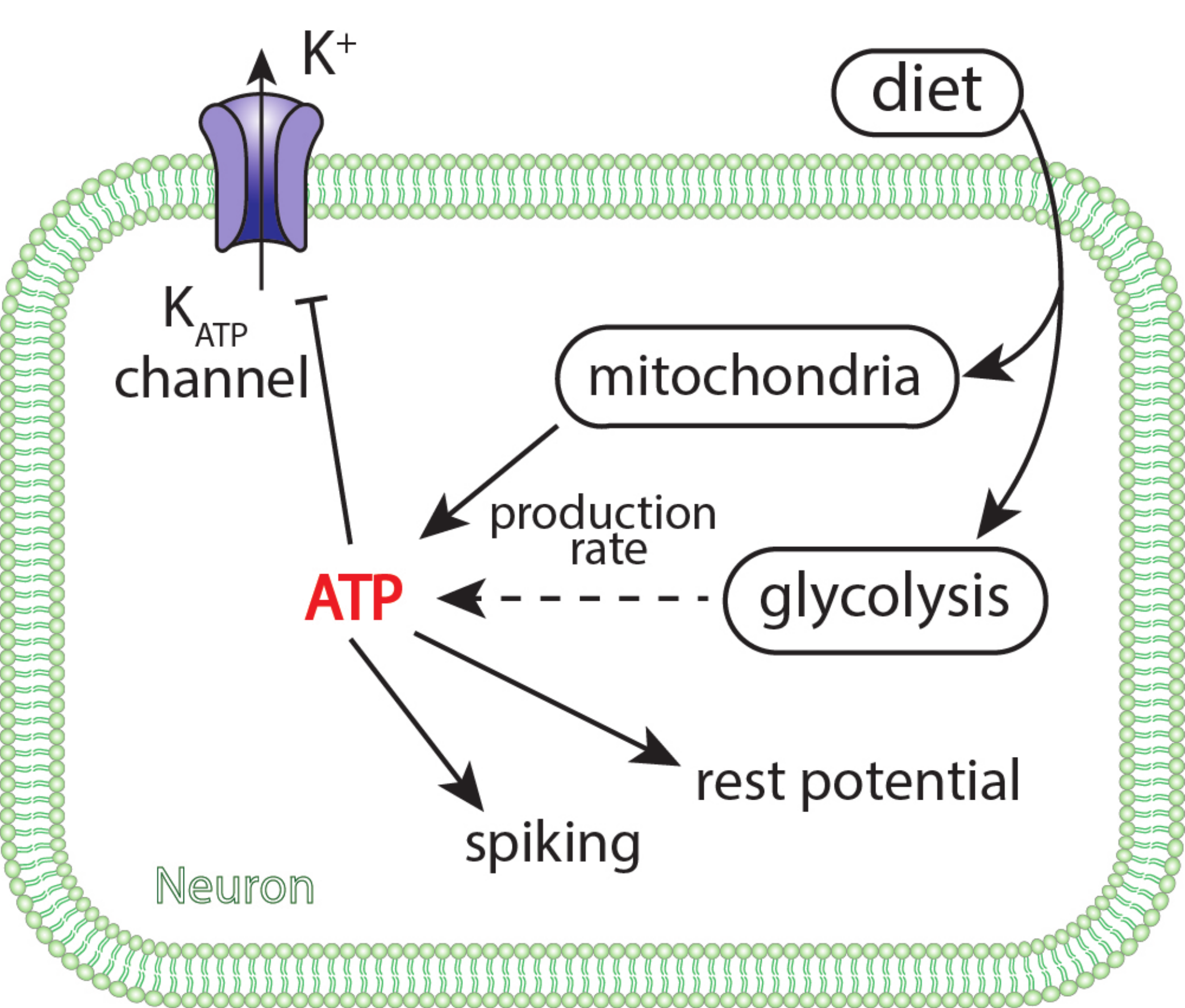}
      \caption{Schematic of the excitability adaptation associated with the changes in the energy metabolism due to ketogenic diet. The diet causes the main mechanism of ATP production to switch from sub-membrane glycolysis to an oxidation process in mitochondria. In parallel, activation of the ATP-gated potassium channels in the neuronal membrane gives rise to an ATP-dependent hyperpolarizing current. ATP balance is further influenced by consumption due to maintaining the 
      resting potential and enabling spiking.} \label{figure1}
\end{figure}

KD is believed to induce two different anti-seizure types of effects \cite{Lutas2013}: one concerns a direct impact of ketone bodies to inhibition of excitatory synaptic channels, and the other is related to the metabolic switch from glycolytic to mitochondria-based ATP production. Our study focuses on the latter group of effects, whose schematic is provided in Fig.~\ref{figure1}. In brief, abolishing the glycolytic ATP production results in the depletion of ATP in a near-membrane region, which causes activation of ATP-sensitive potassium channels in the membrane. Their activation has long been known to induce hyperpolarizing currents that reduce neuronal excitability \cite{Joo2021,Cunningham2006,Ching2012}. Our approach involves a model of local neuronal dynamics supplied with the hyperpolarizing current associated with ATP-gated potassium channels, as well as the ATP dynamics describing its production and utilization.

The paper is organized as follows. In Sec.~\ref{sec:micro}, we introduce the model of a globally coupled heterogeneous population of intrinsically excitable or tonic spiking excitatory quadratic integrate-and-fire neurons supplied by the individual terms for the ATP-dependent hyperpolarizing currents and an equation for the global dynamics of ATP concentration. Using the approach from \cite{Montbrio2015}, in Sec.~\ref{sec:mf} we derive the corresponding mean-field model, which compared to the classical results, contains additional dependencies of the mean firing rate and mean membrane potential on the ATP concentration level. Section \ref{sec:bif} provides the bifurcation analysis of the reduced system, revealing two qualitatively different bifurcation scenarios in terms of transitions to synchrony and the system’s multistability depending on the coupling strength. In Sec.~\ref{sec:switching} we demonstrate mechanisms for controlled switching between the homeostatic and seizure-like states, including external stimulation, parametric and dynamical perturbation. Section \ref{sec:summary} provides a summary and discussion of our results.

\section{Model of microscopic dynamics} \label{sec:micro}

We consider a large heterogeneous population of synaptically coupled excitatory quadratic integrate-and-fire neurons \cite{Ermentrout2010,Izhikevich2007} whose dynamics is given by
\begin{align}
        \dot{V}_j &= V_j^2 +\eta_j + K S(t) + I_j^{\mathrm{ATP}}(t) + I_{\mathrm {ext}}(t) \nonumber \\
        I_j^{\mathrm {ATP}}(t)&= -\alpha V_j(t) \frac{\tilde{C}}{C(t)} \label{eq:QIF}
    \end{align}
where ${V_j}$ for $j=\{1,…,N\},\,N \gg 1$ are the neuronal membrane potentials. The population diversity is manifested through heterogeneity of local bifurcation parameters $\eta_j$, which follow the Lorentzian density distribution 
\begin{align}
g(\eta)= \frac{1}{\pi}\frac{\Delta}{\left(\eta-\bar{\eta}\right)^2+\Delta^2}, \label{eq:gdist}
\end{align}
and may in terms of neurophysiology be interpreted as heterogeneous components of external bias current. Quadratic integrate-and-fire neurons are paradigmatic for type I excitability, such that an isolated unit undergoes a transition from excitable regime ($\eta_j \lesssim 0$) to tonic spiking ($\eta_j >0$) with a frequency $f_j=\sqrt{\eta_j}/\pi$ via a SNIPER (saddle-node of infinite period) bifurcation at $\eta_j=0$. A neuron fires a spike whenever its membrane potential reaches the peak value $V_p$, after which it is reset to $-V_r$. To allow for subsequent analytical tractability, we set $V_p=-V_r=\infty$. 

The terms $I_{\mathrm {ext}}(t)$ and $KS(t)$ describe a common external stimulation and the total synaptic current, respectively. The chemical synapses are characterized by a uniform coupling strength $K$ and are assumed to be instantaneous, such that the mean synaptic activity is given by the normalized output signal of the population 
 \begin{align} 
S(t) = \frac{1}{N}\sum_{j=1}^N\sum_{k\backslash t_j^k \le t} \delta(t-t_j^k), \label{eq:output}
\end{align}
where $t_j^k$ is the time of the k-th spike of neuron $j$.

The excitability adaptation triggered by the changes in the energy metabolism due to the ketogenic diet is mediated by the ATP-dependent hyperpolarizing currents $I_j^{\mathrm{ATP}}$,
which may be regarded as an effective parametric perturbation to the local bifurcation parameters $\eta_j$. The hyperpolarizing currents reflect the action of the ATP-gated potassium channels in neuronal membranes activated by the diet \cite{Joo2021,Huang2007,Martinet2017,Lutas2013,Yellen2008}, and are assumed to have a uniform conductance $\alpha$. Note that the gating of the potassium channels is in fact a complex process governed by the exchange of sodium and ATP \cite{Ching2012,Cunningham2006}. Nevertheless, for simplicity, here we adopt the model for $I_j^{\mathrm {ATP}}$ from \cite{Joo2021} which omits the detailed equation for the sodium dynamics. To close the system ~\eqref{eq:QIF}, one requires an additional equation for the dynamics of ATP concentration $C(t)$
\begin{align}
\dot{C} &=  \frac{\tilde{C}-C}{\tau} - \frac{\epsilon\, S(t)\, C}{\tilde{C}}, \label{eq:atp}
\end{align}
where $\tilde{C}$ denotes the maximal ATP concentration. The first term in the r.h.s. accounts for the ATP production with a characteristic timescale $\tau$, whereas the second term describes the consumption of ATP due to spiking, with $\epsilon$ being the amount of ATP consumed per each spike \cite{Joo2021}. The consumption part is assumed to be proportional to the ATP concentration to keep the concentration from becoming negative. In a notable contrast 
to \cite{Joo2021}, the dynamics of ATP concentration is introduced via a global variable rather than the set of microscopic variables that depend on the spike trains of individual neurons. Adopting such an approach facilitates 
the subsequent derivation of the mean-field model. System ~\eqref{eq:QIF},~\eqref{eq:atp} provides a complete description of a heterogeneous population of quadratic integrate-and-fire neurons subjected to excitability adaptation reflecting the metabolic feedback associated 
with ketogenic diet. 

The difficulties with numerically simulating the reset of membrane potentials in system~\eqref{eq:QIF} are resolved in the standard way by performing simulations of an equivalent model of a population of theta neurons \cite{Ratas2016,Ratas2017}, obtained by the change of variables $V_j=\tan(\theta_j/2)$ \cite{Ermentrout1986,Ermentrout1996} in system~\eqref{eq:QIF}. The phase dynamics of theta neurons then reads
\begin{align}
\dot{\theta}_j&=(1-\cos\theta_j)+(1+\cos\theta_j)(\eta_j+KS(t)+I_{\mathrm {ext}}(t)) \nonumber \\
&-\alpha \frac{\tilde{C}}{C(t)} \sin{\theta_j}, \label{eq:theta}
\end{align}
which should be complemented with Eq.~\eqref{eq:atp} for the dynamics of ATP concentration. The spiking events from the representation ~\eqref{eq:QIF} just correspond to the phase variables’ crossing the $\pi$ value. The Lorentzian distribution of local bifurcation parameters $\eta_j$ is generated deterministically via $\eta_j=\bar{\eta}+\Delta \tan((\pi/2)\frac{2j-N-1}{N+1})$ \cite{Ratas2016}. The numerical simulations of the model ~\eqref{eq:theta},~\eqref{eq:atp} are carried out for the fixed parameter set $\Delta=1.0,\alpha=1.0, \epsilon=1.0, \tilde{C}=1.0, N=10^4$ and varying $\bar{\eta}$, $K$ and $\tau$. While $\bar{\eta}$ and $K$ respectively characterize the intrinsic population dynamics and the impact of interactions in modifying the local dynamics, $\tau$ may 
be seen as the main parameter that describes the metabolic changes associated with the ketogenic diet, consistent with both in-vivo and in-vitro findings that the diet affects the ATP production capacity \cite{Lutas2013,Miller2020,Masino2012,Bough2006,Gano2014,
Saris2022}. For numerical integration of the model ~\eqref{eq:theta} and ~\eqref{eq:atp}, we use an error-controlled 5/4 Runge-Kutta scheme called Tsit5 implemented in the Julia package DifferentialEquations.jl \cite{rackauckas2017differentialequations}. During the simulation, the impact of the spikes and the ATP consumption are handled separately. In particular, we determine $S(t)$ by counting the number of spikes within a given time step and then apply changes to the phases and the ATP concentration according to $\theta_j(t)^+ = \theta_j(t)^- + (1 + \cos(\theta_j(t)^-))KS(t)$ and $ C(t)^+ = C(t)^- - S(t) \epsilon C(t)^- / \tilde{C}$, where the superscripts $-$ and $+$ denote the values before and after the impact of the spikes, respectively. In a subsequent comparison with the 
mean-field model, we present the mean firing rate of the population of theta neurons as a time-smoothed signal of $S(t)$ using a rectangular convolution function with the
time width $0.1$, similar to \cite{Ratas2016}.

\section{Mean-field system}\label{sec:mf}

Applying the general reduction approach from \cite{Montbrio2015} to the microscopic model ~\eqref{eq:QIF}--~\eqref{eq:atp}, in this Section we derive the mean-field system describing the collective behavior of a neuronal population affected by the excitability adaptation reflecting the metabolic changes related to ketogenic diet. In the thermodynamic limit $N \rightarrow \infty$, the population state can be characterized by the conditional density function $\rho(V|\eta,t)$, such that $\rho(V|\eta,t)dV$ gives the fraction of neurons with membrane potential within the interval $(V,V+dV)$ and local bifurcation parameter $\eta$ at time $t$. The density function satisfies the continuity equation
\begin{align}
   \frac{\partial}{\partial t}  \rho &= - \frac{\partial}{\partial V} \left[\rho\left\{ V^2+\eta+KS+I_{\mathrm {ext}} -\alpha V \frac{\tilde{C}}{C} \right\}\right]. \label{eq:cont}
\end{align}

According to the Lorentzian Ansatz \cite{Montbrio2015,Ratas2016,Devalle2018}, connected to the Ott-Antonsen Ansatz via a conformal mapping \cite{Montbrio2015}, the solutions of ~\eqref{eq:cont} generically converge to a Lorentzian distribution \cite{Pietras2023}
\begin{align}
    \rho(V|\eta,t)&=\frac{1}{\pi}\frac{x(\eta,t)}{\left[V-y(\eta,t)\right]^2+x(\eta,t)^2}, \label{eq:Ansatz}
\end{align}
where the time-dependent parameters $x(\eta,t)$ and $y(\eta,t)$ capture the macroscopic dynamics in a reduced subspace. In particular, the local firing rate $r(\eta,t)$ for a fixed $\eta$ is given by the probability flux through the infinity threshold $r(\eta,t)=\rho(V\to \infty|\eta,t)\dot{V}(V\to \infty|\eta,t)$. This yields the relation to the Lorentzian half-width $x(\eta,t)=\pi r(\eta,t)$, such that the population total firing rate satisfies \cite{Montbrio2015}  
\begin{align}
r(t) &= \frac{1}{\pi}\int_{-\infty}^{+\infty} x(\eta,t)g(\eta)d\eta. \label{eq:r}
\end{align}
Similarly, the mean membrane potential is given by 
\begin{align}
v(t) &= \int_{-\infty}^{+\infty} y(\eta,t)g(\eta)d\eta.
\label{eq:v}
\end{align}

Introducing the complex variable $w(\eta,t) = x(\eta,t) + i y(\eta,t)$ and substituting the Lorentzian Ansatz ~\eqref{eq:Ansatz} into the continuity equation leads to
\begin{align}
    \dot{w}(\eta,t)& = i \left(\eta +K r(t)-w(\eta,t)^2+I_{\mathrm {ext}}(t)\right)-\frac{\tilde{C} \alpha w(\eta,t)}{C(t)}, \label{eq:wdot}
\end{align}
where the total output of the finite population $S(t)$ is replaced with the mean rate $r(t)$, the population’s output in the thermodynamic limit. The infinite set of integro-differential equations ~\eqref{eq:wdot} holds for arbitrary choice of the diversity distribution. Nevertheless, the particular choice of Lorentzian distribution ~\eqref{eq:gdist} allows for the maximal reduction of ~\eqref{eq:wdot} by solving the integrals over $\eta$ in Eq. ~\eqref{eq:r} and Eq. ~\eqref{eq:v} via the residue theorem \cite{Montbrio2015,Ratas2016,Ratas2017,Pietras2019,Montbrio2020}. By analytically continuing $\eta$ into the complex plane and closing the integration contour over the half-plane $\operatorname{Im}(\eta)<0$, one readily finds that the integrals in definitions ~\eqref{eq:r} and ~\eqref{eq:v} only depend on the value of $w$ at the pole $\eta=\bar{\eta}- i\Delta$ of the $g(\eta)$ diversity distribution, yielding the relation 
$\pi r(t) + i v(t)=w(\bar{\eta}-i\Delta,t)$. Evaluating Eq. ~\eqref{eq:wdot} at $\eta=\bar{\eta}-i\Delta$ and recalling Eq. ~\eqref{eq:atp}, one arrives at the mean-field model describing the macroscopic dynamics of the population in terms of the mean firing rate $r(t)$, mean membrane potential $v(t)$ and the total ATP concentration $C(t)$
\begin{align}
    \dot{r} &= \Delta/\pi  + (2v-\alpha\tilde{C}/C)r, \nonumber \\
    \dot{v} &=  \bar{\eta} -\pi^2r^2 + v^2+ Kr  -\alpha v\tilde{C}/C + I_{\mathrm {ext}}(t),\nonumber \\
    \dot{C} &=  \frac{\tilde{C}-C}{\tau} - \frac{\epsilon\, r\, C}{\tilde{C}}. \label{eq:mfm}
\end{align}
Note that the system ~\eqref{eq:mfm} is expected to feature a certain separation of timescales, since the ATP production, characterized by the rate $1/\tau$, should take place on a timescale slower than that of the local neuronal dynamics. Nevertheless, there is no reason to a priori assume a strong separation of timescales $1/\tau \ll 1$, so in the following we consider $\tau$ as a free control parameter. One should however bear in mind that the slow-fast decomposition within the NGNM framework has so far provided important insights concerning the onset of collective bursting in neuronal populations with short-term plasticity \cite{Taher2022} and the mechanism behind emergent excitability \cite{Avitabile2022}.

By carrying out the stability and bifurcation analysis of the mean-field model ~\eqref{eq:mfm}, in Sec. IV we demonstrate how the classical physical picture from \cite{Montbrio2015} qualitatively changes in the presence of excitability adaptation associated with ketogenic diet.

\section{Bifurcation analysis of the mean-field model} \label{sec:bif}

\begin{figure*}
  \centering
    \includegraphics[scale=.18]{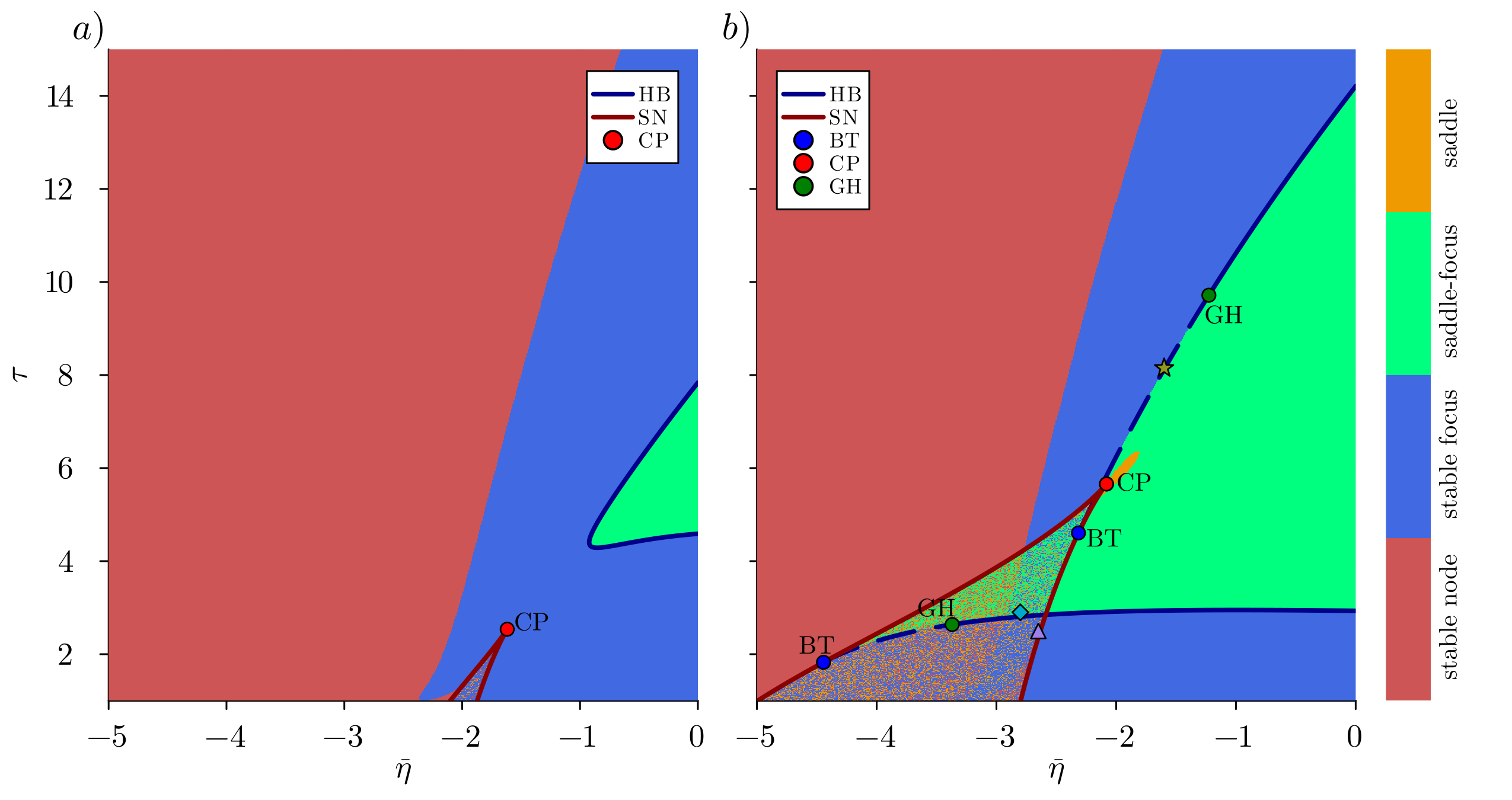}
      \caption{Stability and bifurcations of equilibria of ~\eqref{eq:mfm} in the $(\bar{\eta},\tau)$-plane for coupling strengths $K=10$ and $K=15$ in a) and b), respectively. Shading in different colors, cf. the colorbar, indicates the type of equilibrium (stable node, stable focus, saddle focus, saddle node) obtained by simulating ~\eqref{eq:mfm} from random initial conditions. Bullets denote co-dimension two bifurcations: Bogdanov-Takens (BT), cusp (CP), and generalized Hopf (GH). Lines indicate Hopf (HB) and saddle-node (SN) bifurcation curves (supercritical HB -- solid blue lines, subcritical HB -- dashed blue lines).
      Other symbols (star, triangle, diamond) respectively indicate parameter sets in Fig.~\ref{figure3}, Fig.~\ref{fig:switching4}, Fig.~\ref{fig:switching2} and Fig.~\ref{fig:switching3}. Remaining parameters are $\Delta=1.0, \epsilon=1.0, \alpha=1.0, \tilde{C}=1.0$.} \label{figure2}
\end{figure*}

The stability of equilibria of the macroscopic system  ~\eqref{eq:mfm} and their bifurcations are analyzed by the numerical path-following technique using the software package BifurcationKit \cite{VEL20}. The bifurcation diagrams are presented in the $(\bar{\eta},\tau)$ parameter plane keeping the remaining parameters fixed. The structure of the bifurcation diagrams qualitatively depends on the coupling strength $K$, with the physical picture gaining complexity for larger $K$. To illustrate this, we provide two examples of bifurcation diagrams, namely for $K=10$ in Fig.~\ref{figure2}(a) and $K=15$ in Fig.~\ref{figure2}(b). On top of the bifurcation diagrams are overlaid the fixed-point solutions of ~\eqref{eq:mfm} obtained numerically starting from random initial conditions. The classification of macroscopic equilibria is indicated by the color-coding scheme. One finds four different types of equilibria, including two stable ones (stable node and stable focus) and two unstable ones (saddle and saddle focus).

\begin{figure}
  \centering
    \includegraphics[scale=0.19]{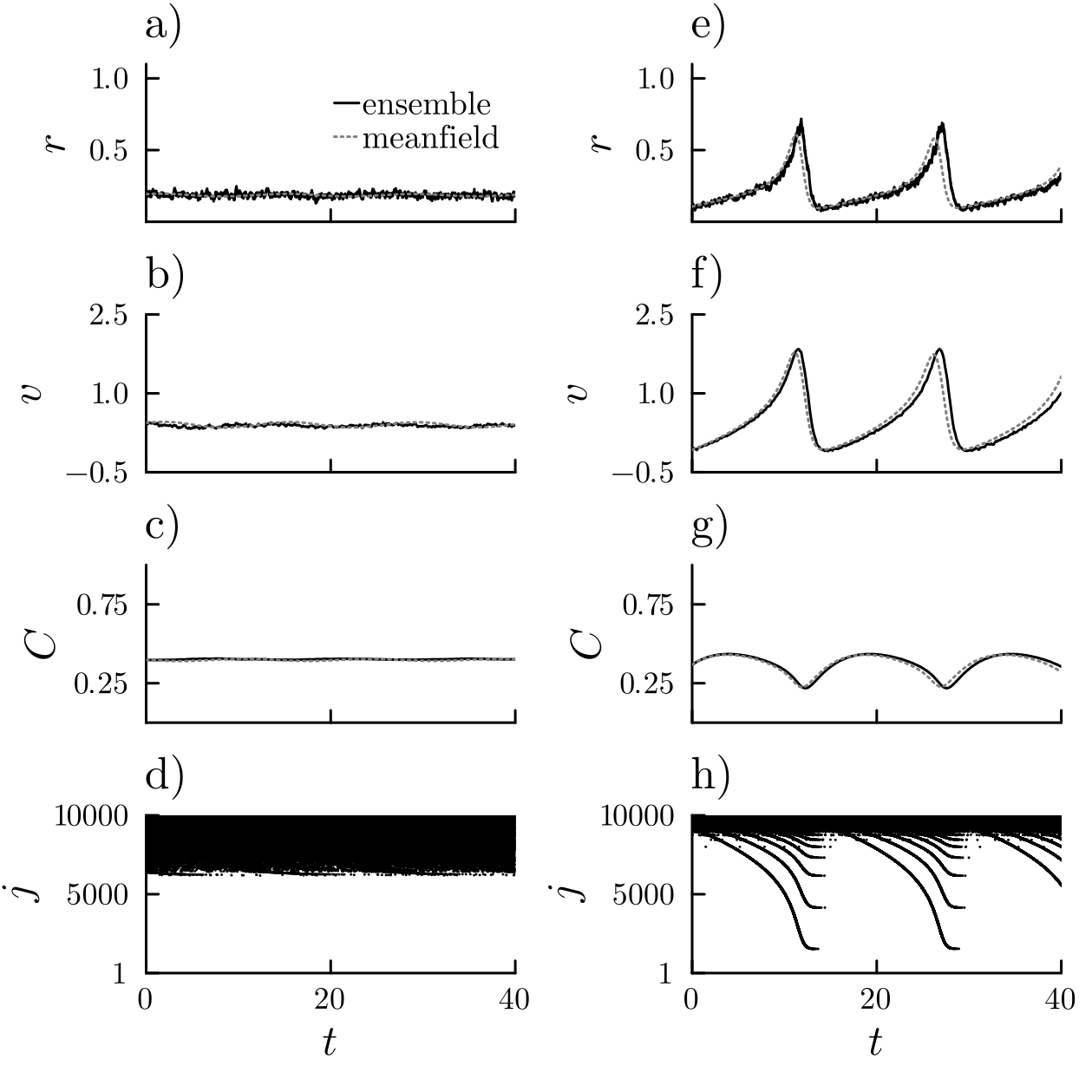}
      \caption{Comparison between the mean-field model~\eqref{eq:mfm} (dashed lines) and microscopic (solid lines) system~\eqref{eq:theta} in terms of $r(t)$, $v(t)$ and $C(t)$ show a good agreement, as well as spike raster plots obtained by simulating the microscopic system for examples of an asynchronous state (a)-(d) and synchronous state (e)-(h). The two states, corresponding to a stable focus and a stable limit cycle of system ~\eqref{eq:mfm}, coexist for $\tau=8.15$, $\bar{\eta}=-1.6$ and $K=15$ (star in Fig.~\ref{figure2}(b)). Remaining parameters: $N=10^4$, $\Delta=\epsilon= \alpha=\tilde{C}=1.0$.}
      \label{figure3}
\end{figure}

The bifurcation diagram for $K=10$ in Fig.~\ref{figure2}(a) features two types of stable equilibria, namely a stable node and a stable focus. The microscopic structure of these states involves asynchronous dynamics and is qualitatively the same for arbitrary $K$, see the example in Figs.~\ref{figure3}(a)-(d) which show the asymptotic dynamics of $r(t)$, $v(t)$, $C(t)$ and the corresponding spike raster plot for the corresponding microscopic model ~\eqref{eq:theta}, respectively. From Eq. ~\eqref{eq:cont}, it follows that for an equilibrium $(r,v,C)=(r^*,v^*,C^*)$, the population is divided into two parts: an excitable (non-spiking) subpopulation for 
$\eta<(\frac{\alpha \tilde{C}}{2C^*})^2-Kr^*$, characterized by the density distribution $\rho(V|\eta)=\delta(V-(\frac{\alpha \tilde{C}}{2C^*}-\sqrt{\frac{\alpha \tilde{C}}{2C^*}^2-(\eta+Kr^*)}))$, and a spiking subpopulation 
$\eta\geq (\frac{\alpha \tilde{C}}{2C^*})^2-Kr^*$ with a Lorentzian-shaped density distribution $\rho(V|\eta)=\frac{\sqrt{\eta+Kr^*-(\alpha \tilde{C}/2C^*)^2}}{\pi((V-\alpha \tilde{C}/(2C^*))^2+\eta+Kr^*-(\alpha \tilde{C}/2C^*)^2)}$. 

The two asynchronous states coexist within a small wedge-shaped region of bistability organized around a co-dimension two cusp point (CP), where two branches of saddle-node bifurcations (SN) meet. Within that region, the stable node (focus) corresponds to the lower (higher)-activity state. Outside of the bistability domain, the mean rate of the nodal equilibrium remains generally low, whereas the focus equilibrium corresponds to the higher-activity state $r^*>0.3$ for sufficiently small $\tau$. From the neuroscience perspective, the lower-activity state indicates the normal, i.e. homeostatic activity of the population. The saddle-node curves for smaller (larger) $\bar{\eta}$ describe the disappearance of the higher-activity (lower-activity) state. 

In contrast to the classical result for an all-to-all synaptically coupled population of quadratic integrate-and-fire neurons \cite{Montbrio2015}, in the presence of excitability adaptation one finds a transition to synchrony mediated by the supercritical Hopf bifurcation (HB), which destabilizes the higher-activity state, cf. the boundaries of the green region in Fig.~\ref{figure2}(a). However, unlike the scenario in mixed populations of excitable units and oscillators with a Kuramoto-type coupling \cite{Klinshov2019}, the branch of Hopf bifurcations does not emanate from a Bogdanov-Takens point. Sufficiently above the Hopf bifurcation, the local dynamics of the synchronous state consists of synchronous population bursts. Within the macroscopic oscillation cycle, one observes an accumulation of synchrony where more units are recruited to emit synchronous spikes as the maxima of $r(t)$ and $v(t)$ are approached, see the example in Figs.~\ref{figure3}(e)-(h). In the neuroscience context, the states with an increased level of synchrony, such as this one, may be interpreted as describing seizure-like dynamics. In particular, in our model, each maximal synchronous burst can be associated with a seizure event. Note that for decreasing adaptation rate (increasing $\tau$), one finds a reentrant transition to  asynchronous state, in the sense that for higher $\tau$ at fixed $\bar{\eta}$, the focus equilibrium regains stability via an inverse supercritical Hopf bifurcation. 

The physical picture substantially changes for a larger coupling strength $K=15$, see Fig.~\ref{figure2}(b). In particular, the bifurcation diagram is organized around seven co-dimension two bifurcation points, including a cusp and pairs of Bogdanov-Takens points (BT), generalized Hopf bifurcations (GH) and fold-homoclinic points (not shown). The most important novelty is that in the presence of excitability adaptation and for sufficiently large $K$, both the lower-activity and the higher-activity asynchronous state may undergo a transition to synchrony. In contrast to Fig.~\ref{figure2}(a), these transitions are associated with branches of Hopf bifurcation curves emanating from the Bogdanov-Takens points. Unlike the classical scenario described in \cite{Ratas2016,Ceni2020}, the Hopf curves exhibit a change of criticality in generalized Hopf (Bautin) points, turning from subcritical to supercritical. From the GH points derive curves of folds of limit cycles, which are not explicitly shown due to additional structures associated with them. Nevertheless, for both GH points, we have been able to numerically corroborate the coexistence between the respective stable equilibrium and the stable limit cycle just above the subcritical Hopf bifurcations. In the neuroscience context, the coexistence between asynchronous and synchronous states has interesting implications in the context of potential control of seizure dynamics, as elaborated in Sec. \ref{sec:switching}.  

Within the wedge-shaped region organized around the CP, the bistability domains between asynchronous and synchronous states are confined by the branches of homoclinic bifurcations. There, the limit cycles derived from HBs associated with the BTs vanish. Nevertheless, the homoclinic bifurcation curves are not shown to avoid overburdening an already complicated bifurcation diagram. We just mention that the homoclinic bifurcation curve associated with the disappearance of the limit cycle born from the destabilization of the higher-activity asynchronous state extends toward the right curve of saddle-nodes for smaller $\tau$. On the other hand, the homoclinic bifurcation associated with the vanishing of the limit cycle derived from the destabilization of the lower-activity asynchronous state remains close to the cusp point, terminating at a fold-homoclinic point on the left branch of the saddle-nodes near the CP. Apart from inside the wedge-shaped region, bistability between asynchronous and synchronous states is also found immediately above the right (higher $\bar{\eta}$) branch of subcritical Hopf bifurcations,
cf. the right dashed line in Fig.~\ref{figure2}(b).

We believe that the qualitative change of the bifurcation diagram observed for varying $K$ is due to the system being close to a co-dimension three bifurcation point. In particular, we suspect that for increasing $K$, there is a scenario of unfolding of a degenerate BT bifurcation, after which the supercritical Hopf curve and the bistability tongue detach from each other. Consequently, there should exist a critical coupling strength, where the two BT points coincide and subsequently disappear. 

In Sec. \ref{sec:switching}, we demonstrate the different strategies that may be applied to induce controlled switches between the seizure-like and homeostatic dynamical regimes.

\section{Control of switching dynamics} \label{sec:switching}

Developing control strategies to suppress the seizure-like states and promote the homeostatic ones is of fundamental importance to potential applications in neuroscience. In Sec. \ref{sec:bif}, we have shown that the presence of excitability adaptation associated with the metabolic changes triggered by ketogenic diet qualitatively enriches the bifurcation diagram compared to the classical scenarios \cite{Montbrio2015,Ratas2016} where such adaptation mechanism is absent. This naturally raises the question of whether the diet-related emergent dynamics may facilitate new means to promote the lower-activity asynchronous states at the expense of regimes featuring excessive synchrony. 

It turns out that there are indeed several control strategies available to induce the desired regime switches. These strategies may be cast into two groups: one associated with inducing critical transitions, and the other related to triggering switches between coexisting states away from criticality. In particular, a supercritical or a subcritical (hysteretic) transition may be induced by the parametric perturbation of the ATP production rate, i.e. the production time constant $\tau$. On the other hand, within the domains supporting bistability between the synchronous and asynchronous states, one may introduce an external stimulation current or apply a dynamic perturbation to ATP concentration level to induce the desired switch of states.

Let us first consider the scenarios concerning the variation of ATP production rate. For lower $K$, the bifurcation diagram in Fig.~\ref{figure2}(a) features a confined synchrony domain (green region) bounded by a reentrant transition to the asynchronous state. In the context of neuroscience, the fact that the transition is mediated by the supercritical Hopf bifurcation implies less sharp onset or termination of seizure-like states. In order to induce a critical transition from synchrony to the asynchronous state, one is required to systematically increase $\tau$, i.e. decrease the ATP production rate, which could in principle be achieved by modulating any number of steps along the metabolic pathway, including the carbohydrate source via the ketogenic diet. Nevertheless, given the reentrant character of the transition, the same effect can be achieved by appropriately reducing $\tau$. In terms of neurophysiology, the former method is still preferred, because the mean rate of the asynchronous state decreases with $\tau$, taking rather low values $r^*\approx 0.2$ immediately above the state’s reappearance.

\begin{figure}[!ht]
  \centering
    \includegraphics[scale=0.15]{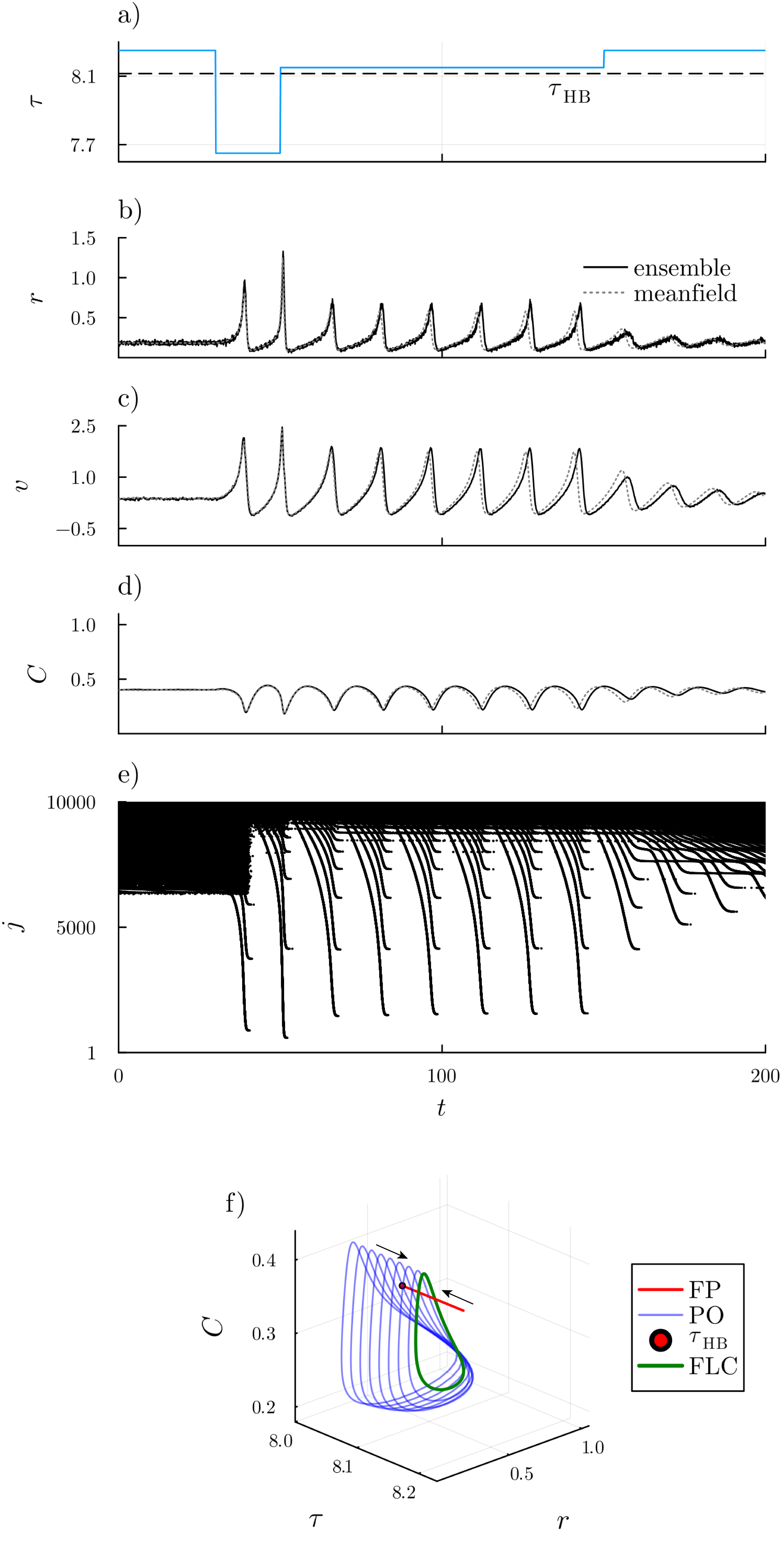}
      \caption{Hysteretic transition between the lower-activity asynchronous state and the synchronous state triggered by sudden changes in ATP production rate in vicinity of subcritical Hopf bifurcation of system ~\eqref{eq:mfm} at $\tau_{\mathrm {HB}}\approx 8.122$. (a) variation of ATP production constant $\tau$; (b)-(d) evolution of macroscopic variables $r(t)$, $v(t)$ and $C(t)$ from~\eqref{eq:mfm} (dashed lines) and their counterparts from the microscopic system ~\eqref{eq:theta} (solid lines); (e) raster plot of spike times for population of $N=10^4$ theta neurons, and (f) system's orbit in phase space of macroscopic variables. Slightly above $\tau_{\mathrm {HB}}$, there is bistability between synchronous and asynchronous states, while below $\tau_{\mathrm {HB}}$ the limit cycle associated with synchronous state is the only attractor. Parameters: $K=15.0$, $\bar{\eta}={-}1.6$, $\Delta=\epsilon= \alpha=\tilde{C}=1.0$, and $N=10^4$.} \label{fig:switching1}
\end{figure}

For the case of stronger couplings from Fig.~\ref{figure2}(b), varying $\tau$ may induce a hysteretic transition in the vicinity of subcritical Hopf bifurcations associated with the destabilization of the lower- or the higher-activity asynchronous state. As an example, let us focus on the case of subcritical Hopf bifurcation controlled by reducing $\tau$ for fixed $K=15$ and $\bar{\eta}=-1.6$. Slightly above the bifurcation, the mean-field system features bistability between the focus equilibrium and the limit cycle illustrated in Fig.~\ref{figure3} for $\tau=8.15$, whereas below the bifurcation, the limit cycle remains the system's only attractor. In Fig.~\ref{fig:switching1}, it is demonstrated how sudden shifts in $\tau$ may induce a hysteretic transition between the asynchronous and synchronous states. In the context of neuroscience, a hysteretic transition between the homeostatic and seizure-like states implies the possibility of a strong jump in the signal amplitude for the seizure onset or termination. Note that a strong jump associated with the hysteretic transition in general implies irreversibility, in the sense that once the jump has occurred, a small reverse of the parameter value cannot induce the transition back to the initial state. In terms of controlling the seizure-like activity, such irreversibility may be favorable because the metabolic interventions that recover the homeostatic state could be persistent and have a lasting therapeutic effect.

The parametric perturbation of the ATP production rate, or rather the ATP production time constant $\tau$, is introduced as follows. Starting from the asynchronous state, a sudden decrease of $\tau$ to the value $\tau=7.65$ below the subcritical Hopf bifurcation at $\tau=\tau_{\mathrm {HB}}$, cf. the dashed line in Fig.~\ref{fig:switching1}(a), triggers the switch to the synchronous state. Then, a sudden increase of $\tau$ to the value slightly above $\tau_{\mathrm {HB}}$ is insufficient to induce a regime shift, i.e. the seizure-like state persists. To suppress the synchronous activity and reinstate the asynchronous state, one is required to introduce an additional parameter perturbation such that $\tau$ exceeds the fold of cycles bifurcation that gives rise to the system's bistability above $\tau_{\mathrm {HB}}$. One observes a good matching between the time series of macroscopic variables $r(t)$, $v(t)$ and $C(t)$, indicated by the solid lines in Fig.~\ref{fig:switching1}(b)-(d), and the corresponding time traces (dashed lines) obtained for the population of theta neurons given by the microscopic model ~\eqref{eq:theta}. 

\begin{figure}[!ht]
  \centering
    \includegraphics[scale=0.15]{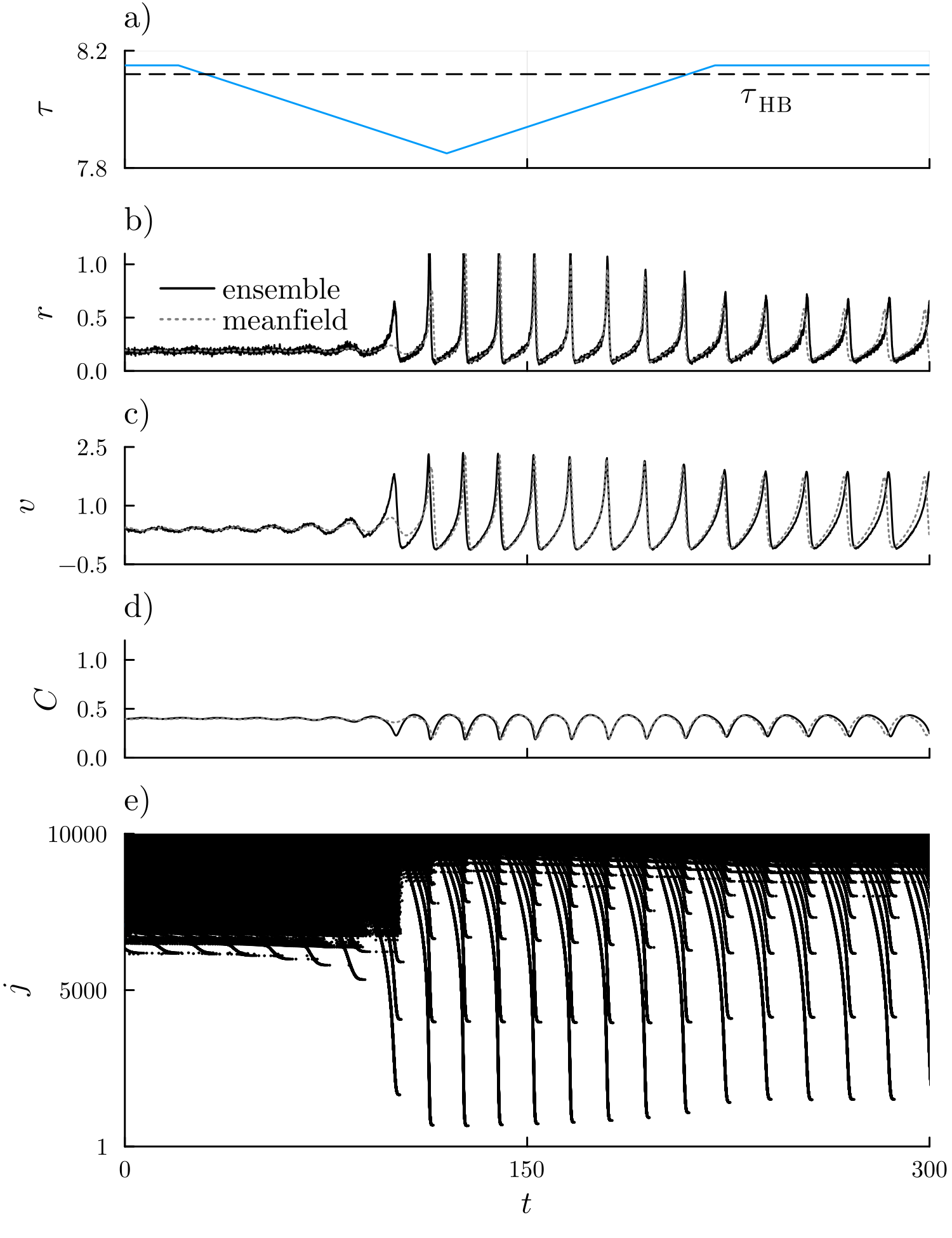}
      \caption{Continuous variation of $\tau$ from $\tau=8.15$ to $\tau=7.85$ and back (a) induces a hysteretic transition between the lower-activity asynchronous state and the synchronous state in the vicinity of the subcritical Hopf bifurcation of system ~\eqref{eq:mfm} at $\tau_{\mathrm {HB}}\approx 8.122$ described in Fig.~\ref{fig:switching1}. Time traces of the macroscopic variables $r(t)$, $v(t)$ and $C(t)$ from~\eqref{eq:mfm} (dashed lines) are compared to their counterparts from the microscopic system~\eqref{eq:theta} (solid lines) in (b)-(d), whereas the spike raster plot for~\eqref{eq:theta}  is shown in (e). One observes a delayed (dynamic) bifurcation to synchronous state as $\tau$ is decreased. Parameters: $K=15.0$, $\bar{\eta}={-}1.6$,
      $\Delta=\epsilon= \alpha=\tilde{C}=1.0$, and $N=10^4$.}
      \label{fig:delayedBif}
\end{figure}

Since implementing sudden parameter changes may not be possible in real experiments, we have also examined a hysteretic transition between asynchronous and synchronous states for a gradual variation in $\tau$, see Fig.~\ref{fig:delayedBif}. The stimulation protocol consists in gradually varying $\tau$ from $\tau=8.15$ to $\tau=7.85$ through the subcritical Hopf bifurcation and back, cf. Fig.~\ref{fig:delayedBif}(a). One observes that the system switches from the asynchronous to the synchronous state and remains there as $\tau$ is increased to its initial value $\tau=8.15$. Moreover, the transition to the synchronous state when decreasing $\tau$ does not occur immediately after reaching the value $\tau=\tau_{\mathrm {HB}}$, which conforms to the effect of delayed (dynamic) bifurcation \cite{Benoit1991,Kuehn2011}, classically observed for slow passages through the bifurcation threshold. While the asymptotic dynamics is captured well by the mean-field model, the transient associated with the delayed bifurcation does not entirely match the behavior of the microscopic system, cf. Figs.~\ref{fig:delayedBif}(b)-(d). Note that the similar hysteretic scenario is observed when considering the subcritical Hopf bifurcation within the wedge-shaped region in Fig.~\ref{figure2}(b).

In the cases when the controlled manipulation of ATP production rate is inaccessible or difficult to achieve, the switches between the seizure-like and homeostatic states may be triggered by introducing appropriate external stimulation currents, or by inducing ATP shocks. The latter conform to the dynamic perturbations of the ATP concentration level, which can be realized in experiments by simply adding or washing out ATP as desired. Both methods can be applied in parameter regions supporting bistability between the synchronous and asynchronous states away from criticality. 

\begin{figure}[!ht]
  \centering
    \includegraphics[scale=0.15]{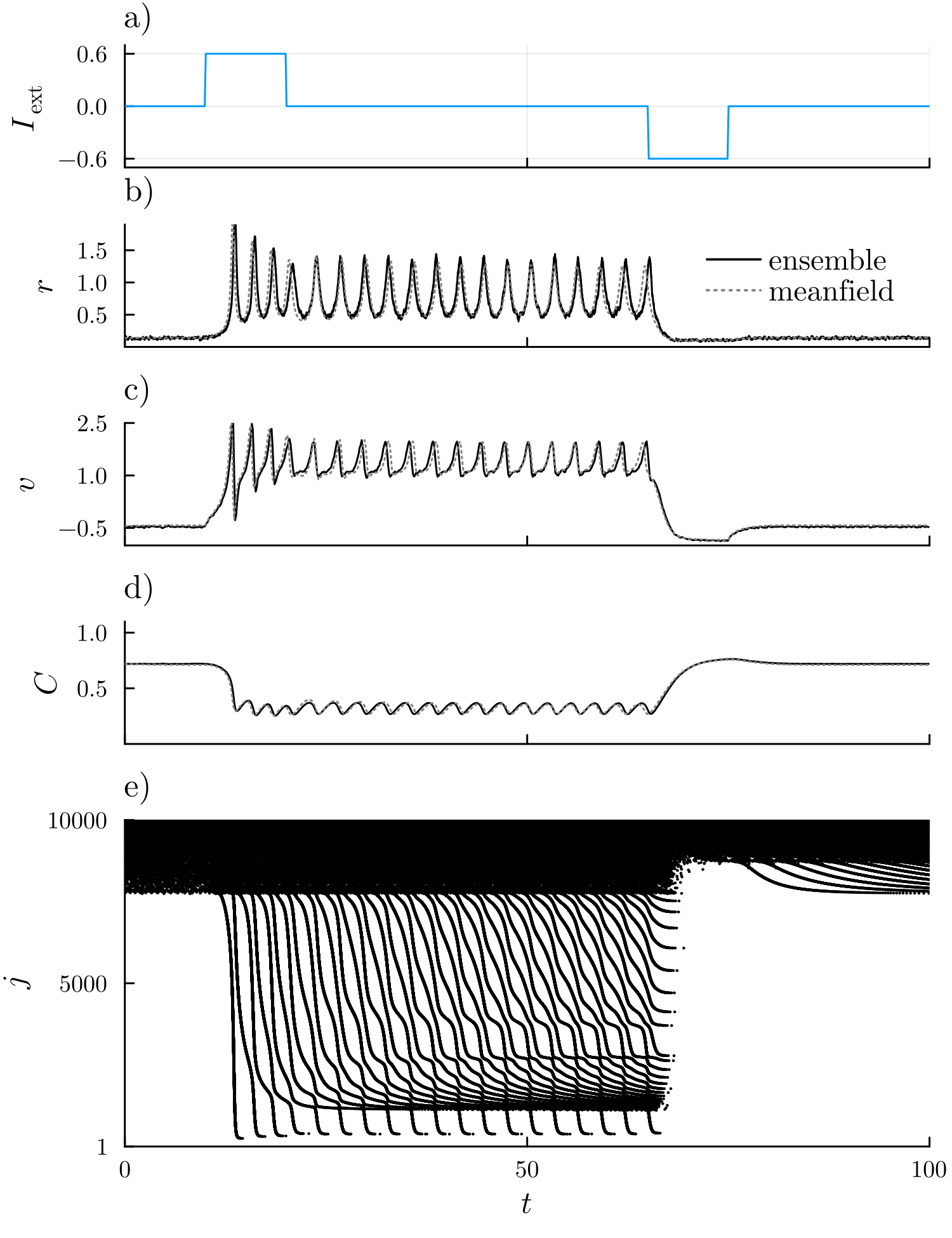}
      \caption{Switch from the lower-activity asynchronous state to the coexisting synchronous state and back induced by excitatory and inhibitory external pulse currents $I_{\mathrm {ext}}(t)$ (a), respectively. (b)-(d) Matching between the time traces of the macroscopic model~\eqref{eq:mfm} and the microscopic system~\eqref{eq:theta} in terms of $r(t)$, $v(t)$ and $C(t)$; (e) Spike raster plot for~\eqref{eq:theta}.  Parameters: $\tau=2.9$, $K=15.0$, $\bar{\eta}={-}2.7$,
      $\Delta=\epsilon= \alpha=\tilde{C}=1.0$ (diamond in Fig.~\ref{figure2}(b)) and $N=10^4$.} \label{fig:switching4}
\end{figure}

In Fig.~\ref{fig:switching4}, it is demonstrated how the excitatory or inhibitory external pulse currents $I_{\mathrm ext}(t)$ may cause the switches in population dynamics. In particular, the inhibitory pulse of sufficient duration can suppress synchrony by inducing the transition to the coexisting asynchronous state. The regime shifts are permanent in the sense that the altered regimes persist after the external stimulation is terminated. Note that the mean-field system captures both the asymptotic and transient dynamics of the microscopic system, see Figs.~\ref{fig:switching4}(b)-(d).

\begin{figure}[!ht]
  \centering
    \includegraphics[scale=0.15]{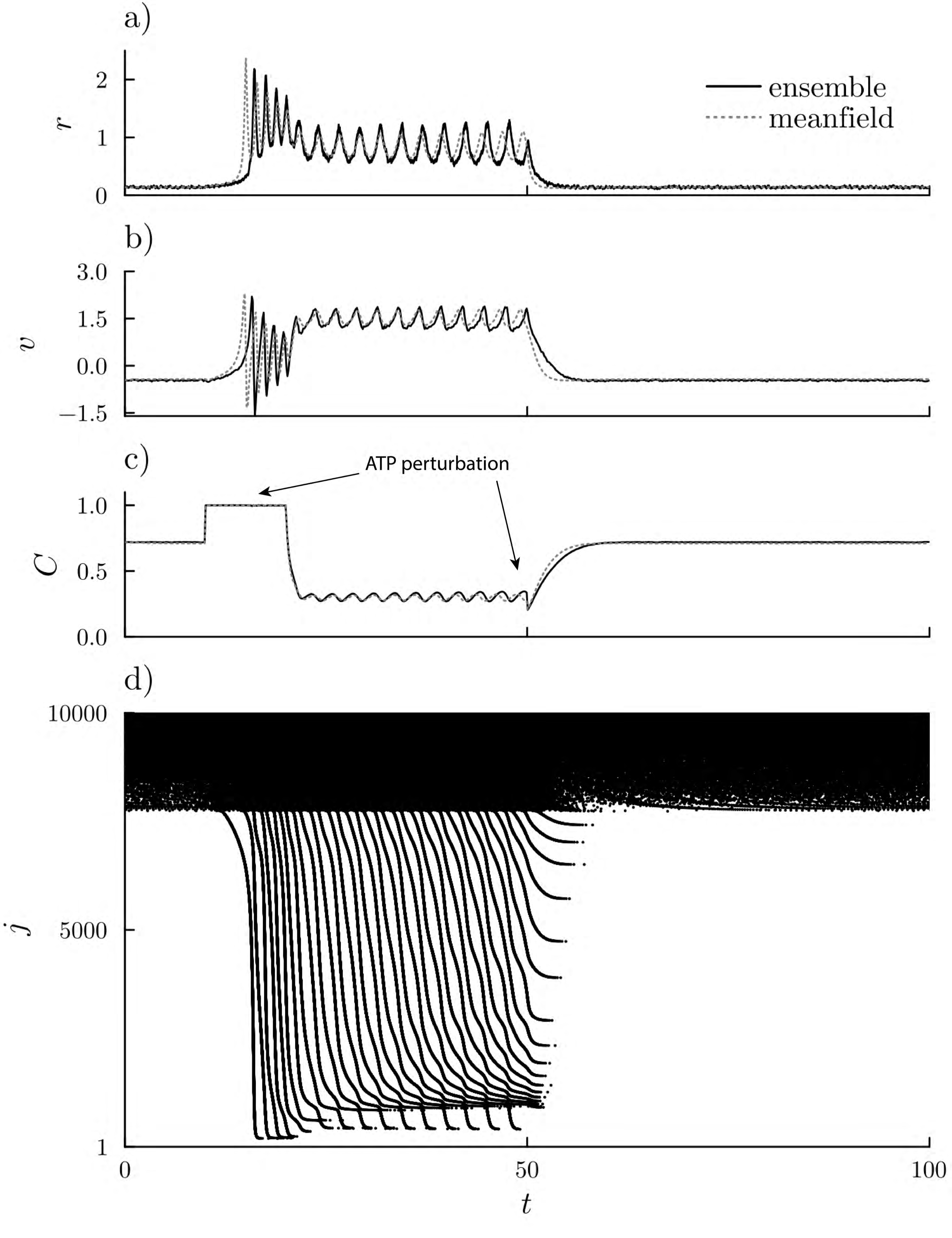}
      \caption{Switches between the coexisting seizure-like and homeostatic states triggered by sudden changes in ATP level, indicated by arrows in (c). A positive ATP shock of finite duration induces the switch from the homeostatic to the seizure-like state. The homeostatic state is regained following an instantaneous, small-amplitude negative ATP shock to the seizure-like state. (a)-(c) Comparison between the time traces $r(t)$, $v(t)$ and $C(t)$ for the mean-field model~\eqref{eq:mfm} (dashed lines) and the corresponding variables of the microscopic system~\eqref{eq:theta} (solid lines); (d) spike raster plot for~\eqref{eq:theta}. Parameters: $\tau=2.9$, $K=15.0$, $\bar{\eta}={-}2.7$,  $\Delta=\epsilon= \alpha=\tilde{C}=1.0$ (diamond in Fig.~\ref{figure2}(b)) and $N=10^4$.}\label{fig:switching2}
\end{figure}

A similar effect can be accomplished by ATP shocks, see Fig.~\ref{fig:switching2}, which comprise sharp externally-induced increases (positive ATP shocks) or decreases (negative ATP shocks) of the ATP concentration level. ATP shocks may permanently alter the system dynamics by switching the system between the attraction basins of the coexisting states. For the
parameters in Fig.~\ref{fig:switching2}, the asynchronous state is the lower-activity equilibrium (stable focus) of system ~\eqref{eq:mfm}, whereas the synchronous state is 
born from the destabilization of the higher-activity asynchronous state. One observes that applying a positive ATP shock of finite width to the asynchronous state is required to trigger the switch to the synchronous state. On the other hand, even an instantaneous negative ATP shock of small amplitude is sufficient to suppress the seizure-like state featuring excessive synchrony and promote the transition back to the homeostatic regime. In principle, the sensitivity of an attractor to switch under dynamical perturbation to a coexisting attractor depends on how close the given attractor lies to the manifold separating between the respective basins of attraction. Note that the mean-field model captures well the impact of the dynamic perturbation associated with ATP shocks, both in terms of transients and the long-term dynamics, see Figs.~\ref{fig:switching2}(a)-(c).

\begin{figure}[!ht]
  \centering
    \includegraphics[scale=0.15]{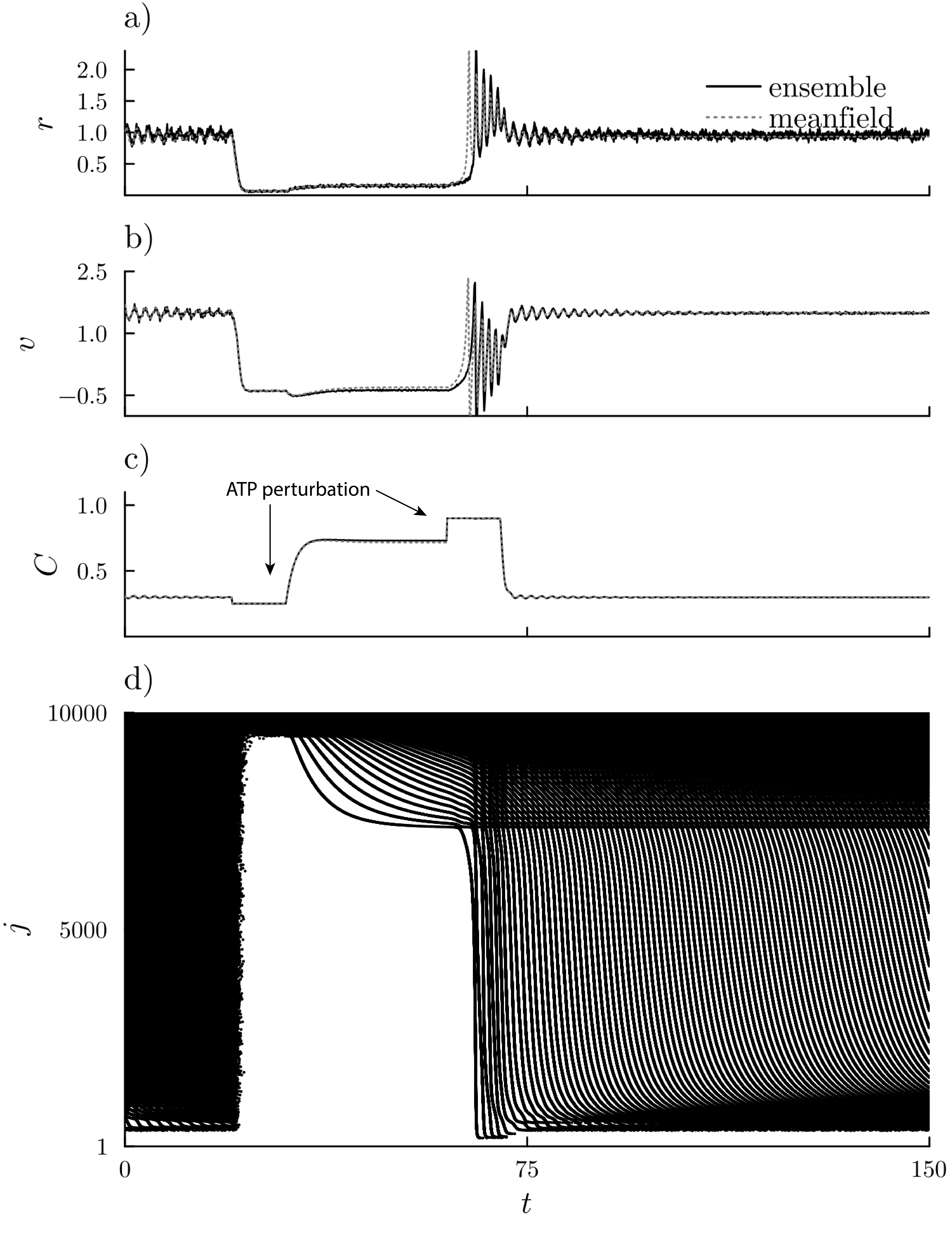}
      \caption{Switch between coexisting higher- and lower-activity asynchronous states (stable foci of system ~\eqref{eq:mfm}) and vice versa triggered by respective negative and positive ATP shocks (see arrows in (c)). (a)-(c) Time traces for the mean-field model $r(t)$, $v(t)$ and $C(t)$ from~\eqref{eq:mfm} (dashed lines) and their counterparts for~\eqref{eq:theta} (solid lines). (d) Spike raster plot for~\eqref{eq:theta}. Parameters: $\tau=2.5$, $K=15.0$, $\bar{\eta}={-}2.65$, $\Delta=\epsilon= \alpha=\tilde{C}=1.0$ (triangle in Fig.~\ref{figure2}(b)) and $N=10^4$.}
      \label{fig:switching3}
\end{figure}

External stimulation currents and ATP shocks can also be applied to trigger switches between the coexisting higher- and lower-activity asynchronous states. An example in Fig.~\ref{fig:switching3} first shows how a brief negative ATP shock induces the regime shift from the higher- to the lower-activity equilibrium. Then, the higher-activity asynchronous state is reinstated by applying a brief positive ATP shock to the lower-activity equilibrium. Note that both the long-term dynamics and the damped oscillations to equilibrium are well described by the mean-field system, see Figs.~\ref{fig:switching3}(a)-(c). 

The examples provided in Sec. \ref{sec:switching} indicate a spectrum of strategies that may in general be applied to regain the homeostatic state, and selecting an optimal strategy in any particular instance is dependent on the corresponding system parameters.  

\section{Summary and discussion} \label{sec:summary}

Within the last two decades, there has been a growing awareness that the feedback of neuronal activity with energy metabolism plays a pivotal role both in maintaining the neuronal homeostasis \cite{Roberts2014,Fields2015,Virkar2016,Kroma2021,Attwell2001,
Franovic2022,Sawicki2023} and in development of certain neurological disorders, such as epilepsy \cite{Fei2020,Rho2017,Meira2019,Rashidy2023} or Parkinson’s disease \cite{Bueller2009,Haddad2015}. Nevertheless, many earlier computational studies have suffered from improperly including the energy constraints, using oversimplified models of local dynamics or implementing too small networks with rather strong finite-size effects. The recent theoretical breakthrough associated with the advent of NGNM models has provided us with a fascinating ability to address the mechanisms of neurological disorders with a strong metabolic component, and to even gain a foothold in understanding the impact of certain therapeutic procedures. In the present paper, we have introduced and analyzed a model of a heterogeneous, globally coupled population of quadratic integrate-and-fire neurons where the individual excitability is modified via an ATP-sensitive potassium current. Implementing the classical reduction approach from \cite{Montbrio2015}, we have derived a three-dimensional mean-field model of population dynamics. The model is intended as a first step toward theoretically explaining the mechanism and apparent effectiveness of ketogenic diet, a long-standing treatment for epilepsy, which nowadays receives a revived interest for helping patients with a drug-resistant form of the disease \cite{Lutas2013,Fei2020}. 

Using the numerical path-following technique, we have analyzed the stability and bifurcations of the derived mean-field model in terms of the ATP production rate and the mean neuronal excitability. The structure of the bifurcation diagram has been shown to qualitatively depend on the coupling strength, gaining complexity for stronger couplings. We have found two qualitatively different types of bifurcation diagrams associated with weaker or stronger couplings. 

For weaker couplings, the diagram features two stable macroscopic equilibria corresponding to the lower- and the higher-activity asynchronous state, which coexist within a small wedge-shaped region bounded by fold bifurcation curves that meet at the cusp. Moreover, for intermediate production rates and a sufficiently high mean excitability, the higher-activity equilibrium undergoes a transition to synchrony via a supercritical Hopf bifurcation. Nevertheless, the asynchronous state is regained for sufficiently small ATP production rates,
i.e. for larger $\tau$. Compared to the classical scenario in heterogeneous populations of quadratic integrate-and-fire neurons with instantaneous chemical synapses \cite{Montbrio2015}, our weak-coupling scenario is richer in the sense that the excitability adaptation gives rise to an additional Kuramoto-like transition to synchrony via supercritical Hopf bifurcation. Nevertheless, it is also simpler than the scenario reported for populations of quadratic integrate-and-fire neurons with finite synaptic time constants \cite{Ratas2016}, populations with instantaneous chemical and electrical synapses \cite{Pietras2019,Montbrio2020}, as well as mixed populations of excitable and oscillatory units with Kuramoto-type interactions \cite{Lafuerza2010,Klinshov2019}, where the transition to synchrony unfolds via the supercritical Hopf bifurcation emanating from a co-dimension two Bogdanov-Takens point. 

For stronger couplings, we have revealed a highly complex bifurcation scenario organized around seven co-dimension two points, namely a cusp and pairs of Bogdanov-Takens, generalized Hopf and fold-homoclinic points. In contrast to \cite{Ratas2016,Pietras2019,Montbrio2020,Lafuerza2010,
Klinshov2019}, the presence of excitability adaptation here facilitates a transition to synchrony not only from the higher-activity asynchronous state but also from the macroscopic equilibrium corresponding to the lower-activity asynchronous state. Moreover, in both instances, the Hopf bifurcations emanating from the Bogdanov-Takens point are subcritical and change character in generalized Hopf points for higher mean population excitability. Remarkably, subcritical transitions to collective oscillations have so far not been observed in massively coupled populations of quadratic integrate-and-fire neurons with only excitatory couplings. Such behavior has rather been found in models involving some form of inhibition and/or connection sparseness \cite{Pyragas2021,Bi2020,Ostojic2009}, or stochastic stimulation \cite{Goldobin2021}. In this context, our results indicate that excitability adaptation may be an important additional source of multistability in the collective dynamics of neuronal populations. 

In relation to explaining the impact of the ketogenic diet as a treatment for epilepsy, we have interpreted the macroscopic equilibrium corresponding to the lower-activity asynchronous state as the “normal”, i.e. the desired state, while the synchronous states have been interpreted as pathological, i.e. seizure-like states. Note that this interpretation is different from \cite{Gerster2021}, the only study so far that has considered the onset and propagation of epileptic seizures within the framework of next-generation neural mass models but focuses on the impact of connectivity rather than neuronal metabolism on the emergent dynamics. In that work, the transition to a high-activity asynchronous state was interpreted as a seizure-like event, but this was likely a consequence of the lack of the transition to synchrony in their model. We have unveiled three different stimulation protocols that may induce controlled switches between the seizure-like and homeostatic states. These methods include parametric perturbation of the ATP production rate, the application of an external (excitatory or inhibitory) pulse current, and a brief pulse-like perturbation of ATP concentration (ATP-shock). The first method consists of inducing a critical transition between the seizure-like and the lower-activity asynchronous state, whereas the latter two methods apply for parameter domains where the population is bistable.  The effectiveness of all three methods is corroborated both for the switching scenarios from the seizure-like to the normal state and back. Interestingly, earlier numerical studies on populations of Hodgkin-Huxley neurons including a similar model of ATP-gated potassium adaptation currents have revealed only a smooth (supercritical) transition from the synchronous (seizure-like) to the asynchronous (normal) state with the increase of ATP production rate \cite{Cunningham2006,Ching2012,Joo2021}. Nevertheless, our study has also revealed the possibility of hysteretic transitions facilitated by the bistability of these states in the vicinity of subcritical Hopf bifurcations, which evinces yet another advantage of introducing the theoretical framework of NGNM models. Note that the irreversible character of hysteretic transitions may prove advantageous for controlling the seizure-like states, given that the induced transition to the lower-activity asynchronous state may then have a persistent therapeutic effect.

Viewed from a different perspective, the possibility of featuring either a continuous or hysteretic transition to seizure-like states may have implications with regard to the ability to predict the onset of seizures from the electroencephalography recordings. Classically, for the continuous transitions, one expects standard indicators of criticality related to critical slowing down \cite{Kuehn2011,Kuehn2013,Meisel2015} to apply, whereas such methods may not hold for hysteretic transitions. In light of the recently sparked controversy as to whether signatures of critical slowing down can indeed be observed prior to epileptic seizures, with arguments provided both pro \cite{Maturana2020} and contra \cite{Wilkat2019}, one may speculate that certain seizures may in fact derive from the continuous, and some from the discontinuous hysteretic transitions, as indicated by our model.

One should point out that the current model does not provide an exhaustive description of the effects of epilepsy and ketogenic diet on the local neuronal kinetics. In particular, epilepsy is known to impact the dynamics of leaky sodium, potassium and chloride currents by triggering changes in the corresponding reversal potentials \cite{Cressman2009, Du2016, Ullah2010}. Also, the current model is limited by the fact that it does not consider the reversal potential of the ATP-dependent potassium current and the way in which it is affected by the ketogenic diet. While this is partly related to the lack of conclusive experimental evidence, it is also associated with the fact that these changes likely cannot be considered isolated from the dynamics of the leaky currents, including their implicit dependencies on surrounding glia, whose taking into account would make the model and its subsequent analysis substantially more complex.  

We believe that our results, together with \cite{Gerster2021, Byrne2019}, have set the stage for opening a new direction of research aimed at theoretical understanding of neurological disorders and associated therapeutic procedures by implementing the next-generation neural mass models. This is expected to provide a valuable complement to experimental studies in at least two ways. Firstly, it will allow for a qualitative insight into multistability which may be lacking or not be easily accessible to experiments; and secondly, it may help in developing effective strategies to control the switching dynamics, especially in terms of suppressing the pathological states. Concerning the research on the ketogenic diet, the next step could involve incorporating some other ingredients that may be related to its therapeutic effects. In particular, one may attempt to derive a “compartmental” model of local dynamics that could distinguish between the two different ATP production processes, namely the classical, glycolytic one near the membrane and the diet-related one unfolding in mitochondria. Also, one may attempt to circumvent the limitations of the current model associated with the assumed uniformity of ATP concentration over the population, aiming to accommodate for the neuron-specific ATP concentration variables that depend on the spike trains of individual neurons \cite{Joo2021}. The microscopic system introduced in the latter way would likely show significant deviations from the current mean-field model, similar to those observed in \cite{Taher2020,Gast2021}. Finally, the immediate impact of ketone bodies on inhibition of excitatory synaptic channels could be taken into account as a factor modulating the synaptic efficacy. More generally, an interesting challenge would be to understand how the classical physical picture developed for activity of neuronal populations within the NGNM framework changes by incorporating the different forms of homeostatic plasticity \cite{Zierenberg2018}, which involve negative feedback loops to constrain the network activity within the desired physiological limit. Also, as a next step, it would be important to provide at least a qualitative validation of our theoretical predictions through elaborately designed experiments, which could then lead to further development and refinement of the current model.

\begin{acknowledgments}
We would like to thank Prof. Eckehard Sch\"{o}ll, Dr. Matthias Wolfrum, and Dr. Predrag Janji\'c for useful discussions. I.F. acknowledges the funding from the Institute of Physics Belgrade through grant by the Ministry of Science, Technological Development and Innovation of the Republic of Serbia and the partial support by the ANSO - Alliance of International Science Organizations Collaborative Research Projects and Training Projects, grant number ANSO-CR-PP-2022-05.
\end{acknowledgments}

\section*{Data Availability}
The data that support the findings of this study are available from the corresponding author upon reasonable request.

\bibliography{refs}

\end{document}